\documentclass[preprint,12pt]{elsarticle}
\usepackage{amssymb}
\usepackage{graphicx}% Include figure files
\usepackage{amsmath,latexsym,amssymb}
\journal{Combustion and Flame}
\begin{document}

\newcommand{\EQ}{\begin{equation}}
\newcommand{\EN}{\end{equation}}
\newcommand{\EQA}{\begin{eqnarray}}
\newcommand{\ENA}{\end{eqnarray}}
\newcommand{\eq}[1]{(\ref{#1})}
\newcommand{\Eq}[1]{Eq.~(\ref{#1})}
\newcommand{\Eqs}[2]{Eqs.~(\ref{#1}) and~(\ref{#2})}
\newcommand{\eqs}[2]{(\ref{#1}) and~(\ref{#2})}
\newcommand{\Eqss}[2]{Eqs.~(\ref{#1})--(\ref{#2})}
\newcommand{\eqss}[2]{(\ref{#1})--(\ref{#2})}
\newcommand{\Section}[1]{Sec.\,\ref{#1}}
\newcommand{\Sec}[1]{\S\,\ref{#1}}
\newcommand{\App}[1]{Appendix~\ref{#1}}
\newcommand{\Fig}[1]{Fig.~\ref{#1}}
\newcommand{\Tab}[1]{Table~\ref{#1}}
\newcommand{\Figs}[2]{Figures~\ref{#1} and \ref{#2}}
\newcommand{\Tabs}[2]{Tables~\ref{#1} and \ref{#2}}
\newcommand{\bra}[1]{\langle #1\rangle}
\newcommand{\bbra}[1]{\left\langle #1\right\rangle}
\newcommand{\mean}[1]{\overline #1}
\newcommand{\meanEMF}{\overline{\mbox{\boldmath ${\cal E}$}} {}}
\newcommand{\meanFF}{\overline{\mbox{\boldmath ${\cal F}$}} {}}
\newcommand{\capback}{\eta_{\mathrm{back}}}
\newcommand{\meanB}{\overline{B}}
\newcommand{\meanF}{\overline{\cal F}}
\newcommand{\meanJ}{\overline{J}}
\newcommand{\meanU}{\overline{U}}
\newcommand{\meanT}{\overline{T}}
\newcommand{\meanrho}{\overline{\rho}}
\newcommand{\UU}{{\bm U}}
\newcommand{\UUp}{{\bm U}_{\rm p}}
\newcommand{\Lg}{L_{\rm g}}
\newcommand{\rhop}{\rho_{\rm p}}
\newcommand{\SSS}{{\sf S}}
\newcommand{\SSSS}{\mbox{\boldmath ${\sf S}$} {}}
\newcommand{\meanAA}{\overline{\mbox{\boldmath $A$}}}
\newcommand{\meanBB}{\overline{\mbox{\boldmath $B$}}}
\newcommand{\meanUU}{\overline{\mbox{\boldmath $U$}}}
\newcommand{\meanJJ}{\overline{\mbox{\boldmath $J$}}}
\newcommand{\meanEE}{\overline{\mbox{\boldmath $E$}}}
\newcommand{\meanuu}{\overline{\mbox{\boldmath $u$}}}
\newcommand{\meanAB}{\overline{\mbox{\boldmath $A\cdot B$}}}
\newcommand{\meanAoBo}{\overline{\mbox{\boldmath $A_0\cdot B_0$}}}
\newcommand{\ff}{\bm{f}}
\newcommand{\xx}{\bm{x}}
\newcommand{\kk}{\bm{k}}
\newcommand{\ii}{\mathrm{i}}
\newcommand{\eee}{\bm{e}}
\newcommand{\meanApoBpo}{\overline{\mbox{\boldmath $A'_0\cdot B'_0$}}}
\newcommand{\meanApBp}{\overline{\mbox{\boldmath $A'\cdot B'$}}}
\newcommand{\meanuxB}{\overline{\mbox{\boldmath $\delta u\times \delta B$}}}
\newcommand{\chk}[1]{[{\em check: #1}]}
\newcommand{\p}{\partial}
\newcommand{\xder}[1]{\frac{\partial #1}{\partial x}}
\newcommand{\yder}[1]{\frac{\partial #1}{\partial y}}
\newcommand{\zder}[1]{\frac{\partial #1}{\partial z}}
\newcommand{\xdertwo}[1]{\frac{\partial^2 #1}{\partial x^2}}
\newcommand{\xderj}[2]{\frac{\partial #1}{\partial x_{#2}}}
\newcommand{\timeder}[1]{\frac{\partial #1}{\partial t}}
\newcommand{\bec}[1]{\mbox{\boldmath $ #1$}}
\newcommand{\nab}{\mbox{\boldmath $\nabla$} {}}

\def\Rey{\mbox{\rm Re}}
\def\Nu{\mbox{\rm Nu}}
\def\Bi{\mbox{\rm Bi}}
\def\Pe{\mbox{\rm Pe}}
\def\Sc{\mbox{\rm Sc}}
\def\St{\mbox{\rm St}}

\begin{frontmatter}
\title{Calculation of the Minimum Ignition Energy based on the ignition 
  delay time}
\author[NTNU]{Jens Tarjei Jensen} 
\author[SINTEF]{Nils Erland L.~Haugen}
\author[Helsinki]{Natalia Babkovskaia}

\address[NTNU]{Department of Physics, Norwegian University of Science and Technology, Trondheim, Norway}
\address[SINTEF]{SINTEF Energy Research, NO-7465 Trondheim, Norway}
\address[Helsinki]{Division of Geophysics and Astronomy (P.O. Box 64),
   FI-00014 University of Helsinki, Finland}

\begin{abstract}
  The Minimum Ignition Energy (MIE) of an initially Gaussian temperature
  profile is found both by Direct Numerical Simulations (DNS) and from a
  new novel model. The model is based on solving the
  heat diffusion equation in zero dimensions for a Gaussian velocity
  distribution. The chemistry is taken into account through the ignition 
  delay time, which is required as input to the model. The model results
  reproduce the DNS results very well for the Hydrogen mixture investigated.

  Furthermore, the effect of ignition source dimensionality is explored, 
  and it is shown that 
  for compact ignition kernels there is a strong effect on dimensionality.
  Here, three, two and one dimensional ignition sources represent a spherical 
  kernel, a long spark and an ignition sheet, respectively.
\end{abstract}

\begin{keyword}
  Combustion \sep ignition  \sep numerics
\end{keyword}
\end{frontmatter}

\section{Introduction}
A flammable mixture of fuel and oxidant exposed to a flame will react and
produce heat and combustion products in a very short time. If the same mixture
is not exposed to a significant heat source at any time, the fuel-oxidant mixture
is, however, stable and will not react. For e.g. safety issues it is important
to know the amount of energy required in order to ignite the mixture and
initiate the combustion process. 
It turns out that in addition to temperature and pressure, the Minimum 
Ignition Energy (MIE) for a given mixture depends on at least
three different parameters;
the geometry of the ignition source, the radius of the ignition source $r_s$ and
the deposition period $t_d$.
For deposition periods in the range $t_a<<t_d<<t_c$, the MIE should be
independent on $t_d$ when $t_a=\delta_f/c$ is the acoustic time scale,
$t_c=\alpha/S_L^2$ is the chemical time scale, $\delta_f \sim \alpha/S_L$
is the thickness of the flame, $\alpha$
is the thermal diffusivity, $S_L$ is the laminar flame speed and $c$ is the
speed of sound. 
Several author groups calculated the MIE using different numerical techniques
\cite{kur04,kondo03,kim04,maas} and experimental investigations \cite{MIEhum}.

Depending on the dimensionality of the heat source the MIE will vary 
considerably. 
The initial ignition kernel may have any given profile, depending on the 
heat source. In the current work a Gaussian profile has been chosen.
A one dimensional heat source correspond to heating the mixture in an infinitely
large plane, which could possibly be realized by laser sheets.
A two dimensional heat source correspond to an infinitely long 
cylindrical ignition kernel. This is in essence the same as a spark ignition where
the length of the spark is significantly longer than its width. Finally
a Gaussian profile in three dimensions correspond to a spherical source which
could be thought of as a very short spark,
or as ignition by the combined effect of several focused lasers.

In the present work it is assumed that all thermal energy is deposited 
with a Gaussian distribution and constant pressure prior
to starting the calculation. This essentially means that $t_a<<t_d<<t_c$.
This simplification has been chosen in order to remove one variable from the
equation and consequently to more easily see the
fundamental underlying physics. Furthermore the focus is on Hydrogen, but
the methods and conclusions should qualitatively be independent on
the fuel and therefore be of generic interest.

\section{Numerical simulations}

The minimum ignition energy for a combustible hydrogen mixture is
found be running a series of Direct Numerical Simulations (DNS) with
the Pencil Code\cite{PC,bab11}. 
Unlike Large Eddy
Simulations (LES) and Reynold-average Navier-Stokes simulations
(RANS), which use turbulence modelling, DNS solves the full
Navier-Stokes equations without the use of any modelling
and filtering.

To simulate the physical problem the conservation equations for mass, momentum,
species and energy have to be solved together with the equation of state.
The equation for conservation of mass is given as
\begin{equation}
  \label{EQ: Continuity}
  \frac{D \ln \rho }{D t}
  = -\mathbf{\nabla}\cdot\mathbf{U},
\end{equation}
where $\mathbf{U}$ is the velocity vector, $\rho$ is the density and
$D/Dt = \partial/\partial t + \mathbf{U}\cdot\mathbf{\nabla}$ 
is the advective derivative.
 
The momentum equation has the form
\begin{equation}
  \label{EQ: Momentum}
  \frac{D{\mathbf{U}}}{D{t}}
  = \frac{1}{\rho}(-\mathbf{\nabla}p + \mathbf{F}_{vs}),
\end{equation}
where $\mathbf{F}_{vs}=\nabla \cdot(2\rho\nu\mathbf{S})$ 
is the viscous force,  
$\mathbf{S}$ is the rate of strain tensor,
and $p$ is the pressure. The
species evolution equation is
\begin{equation}
  \label{EQ: Species}
  \frac{D{Y_{j}}}{D{t}}
  = \frac{1}{\rho}\left(-\mathbf{\nabla}\cdot\mathbf{J}_{j} + \dot{\omega_{j}}
\right),
\end{equation}
where $\dot{\omega_{j}}$ is the reaction rate, $\mathbf{J}_{j}$ is the
diffusive flux and $Y_{j}$ is the mass fraction of specie $j$. 
Lastly, the energy equation is solved for the temperature 
\begin{equation}
  \label{EQ: Energy}
  \left(c_{p} - \frac{R}{M}\right)\frac{D\ln T}{Dt} 
  =  \displaystyle\sum\limits_{j}^{} \frac{DY_{j}}{Dt} 
  \left(\frac{R}{M_{j}} - \frac{h_{j}}{T}\right) 
  - \frac{R}{M} \mathbf{\nabla}\cdot\mathbf{U}+\frac{2\nu\mathbf{S}^{2}}{T} 
  - \frac{\mathbf{\nabla}\cdot\mathbf{q}}{\rho T},
\end{equation}
where $c_p$ is the heat capacity at constant pressure, $R$ is the
universal gas constant, $M$ is the molar mass, $T$ is the temperature,
$h$ is the enthalpy and $\mathbf{q}$ is the heat flux.
For temporal discretization the Pencil Code uses a
third order Runge-Kutta method, while a sixth-order central difference
scheme is used for the spatial discretization.
For a more thorough discussion on the equations solved 
see Babkovskaia, Haugen and Brandenburg (2011) \cite{bab11}.

\begin{figure}
  \includegraphics[scale=0.4]{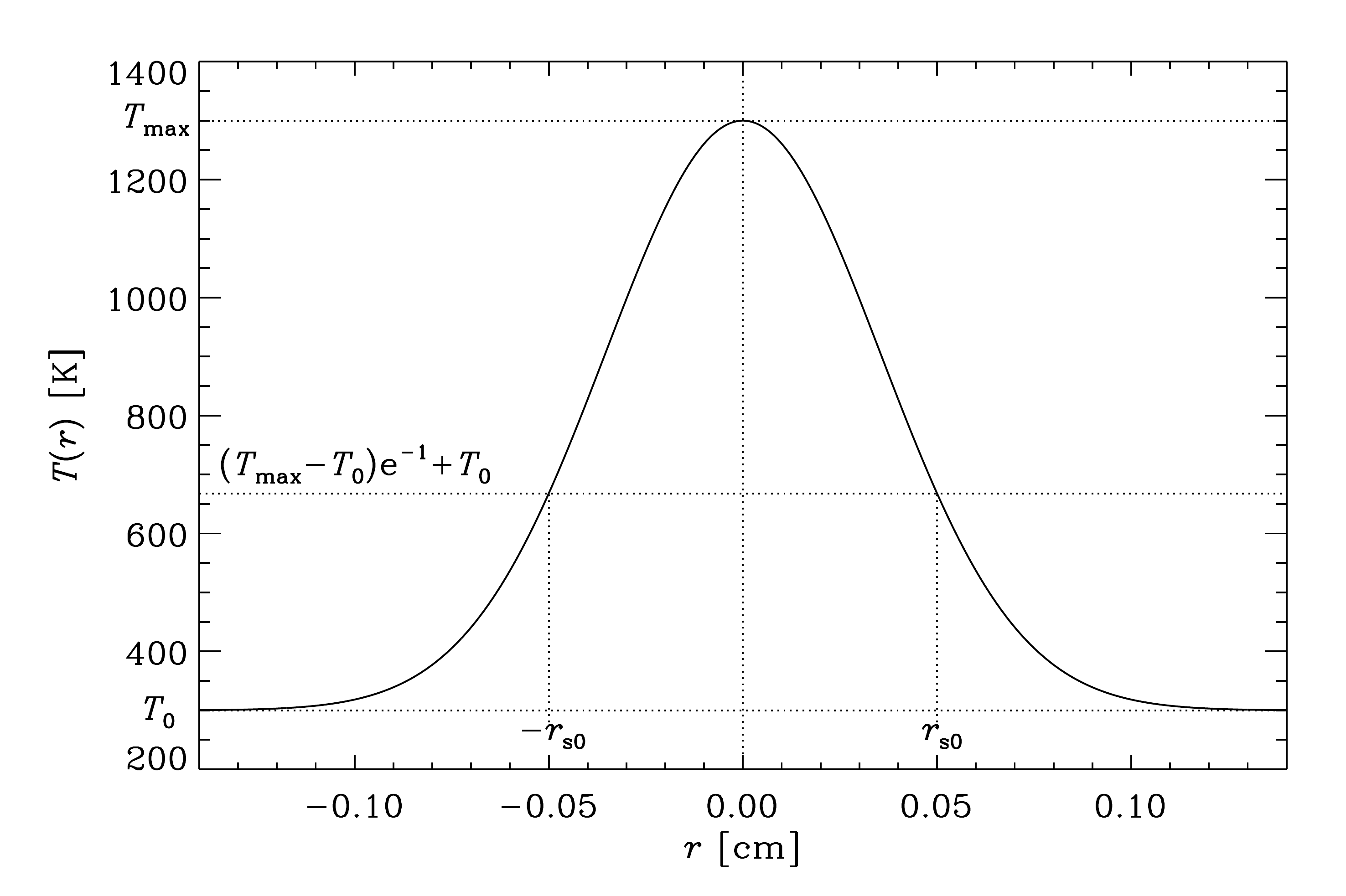}
  \caption{Illustration
    of the initial Gaussian temperature distribution, where $T_0=300$~K,
    $r_\mathrm{s_0}=0.05$~cm, $T_\mathrm{max} = 1300$~K and the size of the domain $L=0.3~\text{cm}$.
    \label{FIG: Gaussian distribution}}
\end{figure}

To simulate an ignition source an initial
temperature distribution is imposed in the domain. The distribution
is chosen to be Gaussian, and parallels could be drawn to a real life
heat source that has its hottest spot in the center. For instance, if
a cross-section of a real electrical spark is made, the temperature
distribution at this cross-section could compare well with a Gaussian
shaped temperature distribution in two dimensions. The initial
distribution is given by 
\begin{equation}
\label{EQ: Gaussian Temp Dist}
T(r) = \left(T_\mathrm{max}-T_{0}\right) e^{-\left(\frac{r}{r_{s_0}}\right)^{2}} + T_{0},
\end{equation}
where $T_\mathrm{max}$ is the maximum temperature, $T_{0}$ is ambient
temperature and $r_\mathrm{s_0}$ is the radius of
the distribution. 
\Fig{FIG: Gaussian distribution} illustrates
the initial temperature profile in one dimension for one of the simulations.

To simulate a closed vessel or container 
the spatial derivative of the temperature, species and the density, together
with the value of all velocity components, are set to zero
at the walls. The
domain is chosen to be sufficiently large, which implies that
all spatial gradients are close to zero at the boundaries
for all relevant times. The size of the domain, $L$, ranges from $0.5~\text{cm}$ for $r_{s0} = 0.1~\text{cm}$ to $0.06~\text{cm}$ for $r_{s0}=0.01~\text{cm}$.

In this work the fuel is chosen to be Hydrogen. This choice is made due to
the relative simplicity of the Hydrogen reaction mechanism. 
The flammable mixture thus consist of dry air mixed with hydrogen,
so the initial species are $H_{2}$, $O_{2}$ and $N_{2}$. 
For all our simulations we take the equivalence ratio of 0.8.
We have
assumed that all the minor species in the air, such as argon and
carbon-dioxide are negligible and that the initial gas is completely
homogeneous.

Radicals are not included in the initial mixture, even though the
temperature profile already exists in the system at time zero. The
radicals will start to form immediately after the simulation is
started.

\section{Model description}

In the following a new, novel, model for calculation of MIE is described.
The model accounts for the
chemistry only through the ignition delay time,
and it only considers the central point in the temperature profile. 
Since a Gaussian temperature profile is assumed for all times,
the spatial gradients are easily found when $T_\mathrm{m}(t)$ and $r_s(t)$
are known.

\subsection{Obtaining an expression for the central temperature}
\label{SECT: Center T}
In this subsections it will be shown how the Gaussian
temperature distribution and the heat equation are used to find
an expression of the central temperature as a function of
time. In later subsections it will be illustrated how this expression is
connected with the ignition delay time, and how the model determines a
case of ignition. 

A large closed volume $V$ is considered.
Given that the initial temperature distribution is Gaussian it is assumed that
the distribution stays Gaussian also during the short period until it can be 
determined if the mixture ignites or not.
When the chemical reactions
start influencing the temperature it is, however, clear that the profile will 
differ from Gaussian as the chemical heating initially will occur
only in the center of the profile.
The temperature distribution is then given by
\begin{equation}
\label{EQ: T(r,t)}
T\left(r,t\right) 
= \left(T_\mathrm{m}(t)-T_{0}\right) e^{-\left(\frac{r}{r_{\mathrm{s}}(t)}\right)^2} 
+ T_{0},
\end{equation}
where $T_{0}$ is the ambient temperature, $T_\mathrm{m}(t)$ is the
temperature in the middle of the distribution
and $r_{\mathrm{s}}(t)$ is the radius of the heat
source, in this case the standard deviation. The initial values at
$t=0$ are defined to be
\begin{align}
T_\mathrm{m}(0)&= T_{\mathrm{max}} \label{EQ: IC Tm(t)}\\
r_{\mathrm{s}}(0)&=r_{\mathrm{s_0}}. \label{EQ: IC rs(t)}
\end{align}

The heat equation is required in order to include heat
diffusion, and is given as
\begin{equation}\label{EQ: heat1}
\frac{\partial{T}}{\partial{t}} = \alpha\nabla ^2 T,
\end{equation}
where $\alpha$ is the thermal diffusivity. In 2D, it is convenient to
use the polar coordinate system, and in 3D it is the spherical
coordinate system which is most convenient. Since both
$\theta$ an $\phi$ are set to be symmetric, the derivatives with
respect to $\theta$ and $\phi$ will be zero in our system. Hence, the
Laplace operator is 
\EQ
\nabla ^2 = \frac{1}{r^{N-1}}\frac{\partial}{\partial r}\left(r^{N-1} \frac{\partial}{\partial r}\right),
\EN
where $N=$ 1, 2 or 3 is the number of dimensions.

By inserting the Gaussian temperature distribution, \Eq{EQ: T(r,t)}, into 
\Eq{EQ: heat1}, and
then evaluate the remaining equation at $r=0$,
it simplifies to
\begin{equation}
\label{EQ: 1.order DE Tm(t) with rs(t)}
\frac{\partial T_\mathrm{m}(t)}{\partial t} = - 2\alpha(T_\mathrm{m}) N\frac{T_\mathrm{m}(t)-T_0}{r_{\mathrm{s}}(t)^2}.
\end{equation}

Since a closed volume $V$ is considered the specific volume $n=V/M_{\rm total}$, 
where $M_{\rm total}$ is the total mass within the volume, must be 
conserved. This yields
\EQ
n-n_0
=\int_V \frac{1}{\rho} - \frac{1}{\rho_0} dv
=\frac{\cal{R}}{P}\int_V T - T_0 dv=\mbox{constant}
\EN
when $n_0$ is the reference specific volume for $T_{\rm max}=T_0$ and $\cal{R}$
is the universal gas constant divided by the mean molar mass.  
Combining this with \Eqss{EQ: T(r,t)}{EQ: IC rs(t)} gives
\begin{equation}
(T_\mathrm{m}(t)-T_{0}) \cdot r_{\mathrm{s}}(t)^N 
= (T_\mathrm{max}-T_{0}) \cdot r_{\mathrm{s_0}}^{N}.
\end{equation}
From this equation $r_s(t)$ can be solved for, such that
\begin{equation}\label{EQ: rs(t) with T(t)}
r_{\mathrm{s}}(t)=r_{\mathrm{s_0}}\bigg(\frac{T_{\mathrm{max}} - T_0}{T_\mathrm{m}(t)-T_0}\bigg)^\frac{1}{N},
\end{equation}
where $N$ again is the number of dimensions.  The variable $r_s(t)$
can now be replaced in \Eq{EQ: 1.order DE Tm(t) with rs(t)} 
to obtain a solvable first order differential equation,
which is given as

\begin{equation}\label{EQ: 1.order DE Tm(t)}
\frac{\partial T_\mathrm{m}(t)}{\partial t} = -2\alpha(T_\mathrm{m}) N \frac{\left(T_\mathrm{m}(t)-T_0\right)^\frac{2+N}{N}}{r_{\mathrm{s_0}}^2 \left(T_{\mathrm{max}} - T_0\right)^\frac{2}{N}}.
\end{equation}

The thermal diffusivity $\alpha(T)$ is here based on empirical thermodynamical 
data and fitted by the polynomial
\begin{equation}
\alpha\left(T_\mathrm{m}(t)\right)=\left(E+DT_\mathrm{m}(t)+CT_\mathrm{m}^2(t)+BT_\mathrm{m}^3(t)+AT_\mathrm{m}^4(t)\right),
\end{equation}
where the parameters are given in Table \ref{TABLE: alpha parameters}.
\begin{table}
\centering
\caption{The parameters from the regression of $\alpha(T)$}
\begin{tabular}{p{3cm} p{3cm}}
  \hline \hline
Parameter & Value\\ 
  \hline \\
A & 0.133608 $\cdot 10^{-13}$ \\
B & 0.208675 $\cdot 10^{-9}$ \\
C & 0.234699 $\cdot 10^{-5}$ \\
D & 0.971533 $\cdot 10^{-3}$ \\
E & 0.257533 $\cdot 10^{-5}$ \\ [1ex] 
  \hline
  \label{TABLE: alpha parameters}
\end{tabular}
\end{table}

To solve \Eq{EQ: 1.order DE Tm(t)}, which numerically is only zero
dimensional, straight forward time stepping is used. The method
applied in this work is the fourth order Runge-Kutta method.

\subsection{The ignition delay time}

The elements included in the model so far do not involve chemistry. To
account for the chemistry involved in an ignition process the ignition
delay time is used.

To be able to predict ignition in a given mixture, one need to obtain
data on the ignition delay time from that specific mixture. 

In order to measure the ignition delay time, a definition of
when an ignition has taken place is required. 
There are several different definitions available in the literature, two
definitions which are often used are 
 1) 
the time until the temperature has increased 400~K past the initial temperature
and, 2) 
the time until the maximum temporal derivative of the temperature is achieved.

\begin{figure}
\includegraphics[scale=0.45]{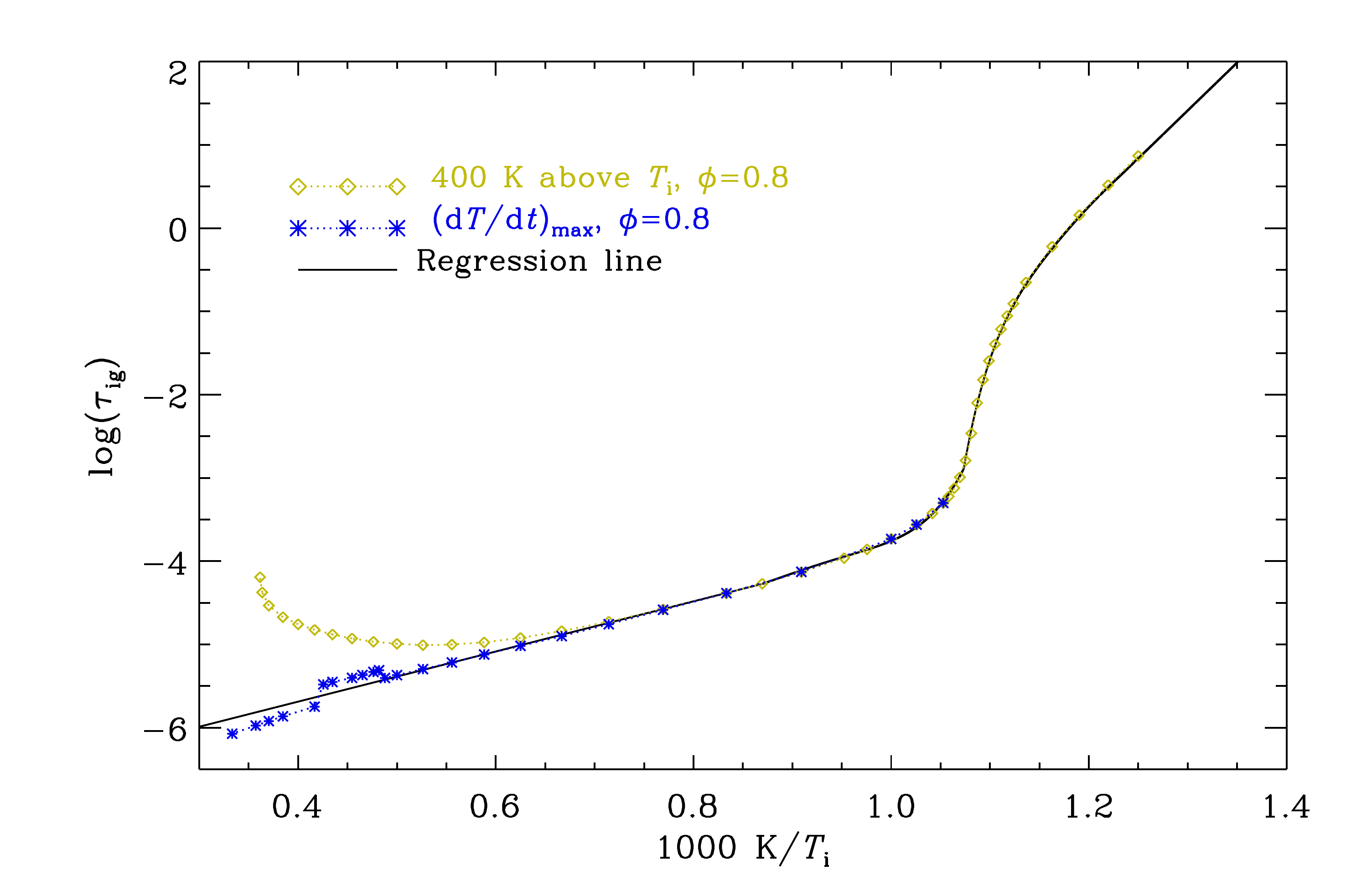}
\caption{Ignition delay time as a function
  of temperature for two different definitions of the ignition delay time.
The yellow line present the results found by defining ignition as the point 
where the temperature has reached 400 degrees above the initial temperature, 
while for the blue line ignition is defines as the time when the temperature
gradient is at its maximum. The black line is a best fit to the blue line.}
\label{FIG: ignDelay_graph}
\end{figure}

\Fig{FIG: ignDelay_graph} shows how the ignition delay time
depends on temperature. In the figure there are two data sets, where both
are obtained from zero dimensional 
simulations with the Pencil Code. The two are based on 
different methods of ignition determination. It can be seen that they differ
when the temperature is sufficiently high. 
This is due to the fact that for high temperatures an increasing amount of the 
chemical energy is converted into radicals instead of thermal energy. This means
that above a certain preheat temperature the mixture temperature 
will not increase with as 
much as 400~K, and this leads to the
very rapid increase of $\tau_\mathrm{ig}$ for definition 1).

The model use $\tau_\mathrm{ig}(T)$ together
with the heat equation to
predict if a mixture ignites or not, 
based on mixture composition and the initial parameters 
$T_\mathrm{max}$ and $r_\mathrm{s_0}$.
Lets define an ignition progress variable $P$ such that $P=0$ at $t=0$ and
$P=1$ when the system has reached ignition. If $P$ does not reach 1 this
means that the mixture did 
not ignite, i.e. not enough energy was supplied to the system.
An equation, which fulfill these requirements, is
\begin{equation}
\label{EQ: Ignition P}
P= \int\limits_{t=0}^{\infty} \! \frac{1}{\tau_\mathrm{ig}(T_\mathrm{m}(t))} \, \mathrm{d}t.
\end{equation}
The rationale behind this equation is that in order for ignition to occur a
certain amount of radicals are required. These radicals are produced at all
temperatures above a certain limit. 
The rate of radical production is inversionally proportional 
to the ignition delay time.

Initially the
parameters $T_\mathrm{max}$ and $r_\mathrm{s_0}$ are given.
For each time step of lenght $\Delta t$, some heat diffuses away, and a
new temperature, $T_\mathrm{m}(t+\Delta t)$, is obtained from 
\Eq{EQ: 1.order DE Tm(t)}. Based on the new temperature a new
ignition delay time, $\tau_\mathrm{ig}\left(T_\mathrm{m}\left(t+\Delta
t\right)\right)$, is obtained, which gives a new contribution to the
ignition progress variable $P$. If, at any given time, P equals unity 
the current parameters and conditions have produced an ignition. 
As seen from \Fig{FIG: ignDelay_graph}, the lower the
temperature gets, the higher the ignition delay time gets. This means
that the term $1/\tau_\mathrm{ig}(T_\mathrm{m}(t))$ in \Eq{EQ: Ignition P} 
gets smaller, and each contribution towards $P$
for each time step gets smaller. If heat is diffused away too quickly
$P$ will never reach 1, and the process will count as a non ignition case.

\section{Results}
In all of the following an hydrogen-air mixture with an equivalence ratio
of 0.8 is considered.
\begin{figure}
\includegraphics[scale=0.4]{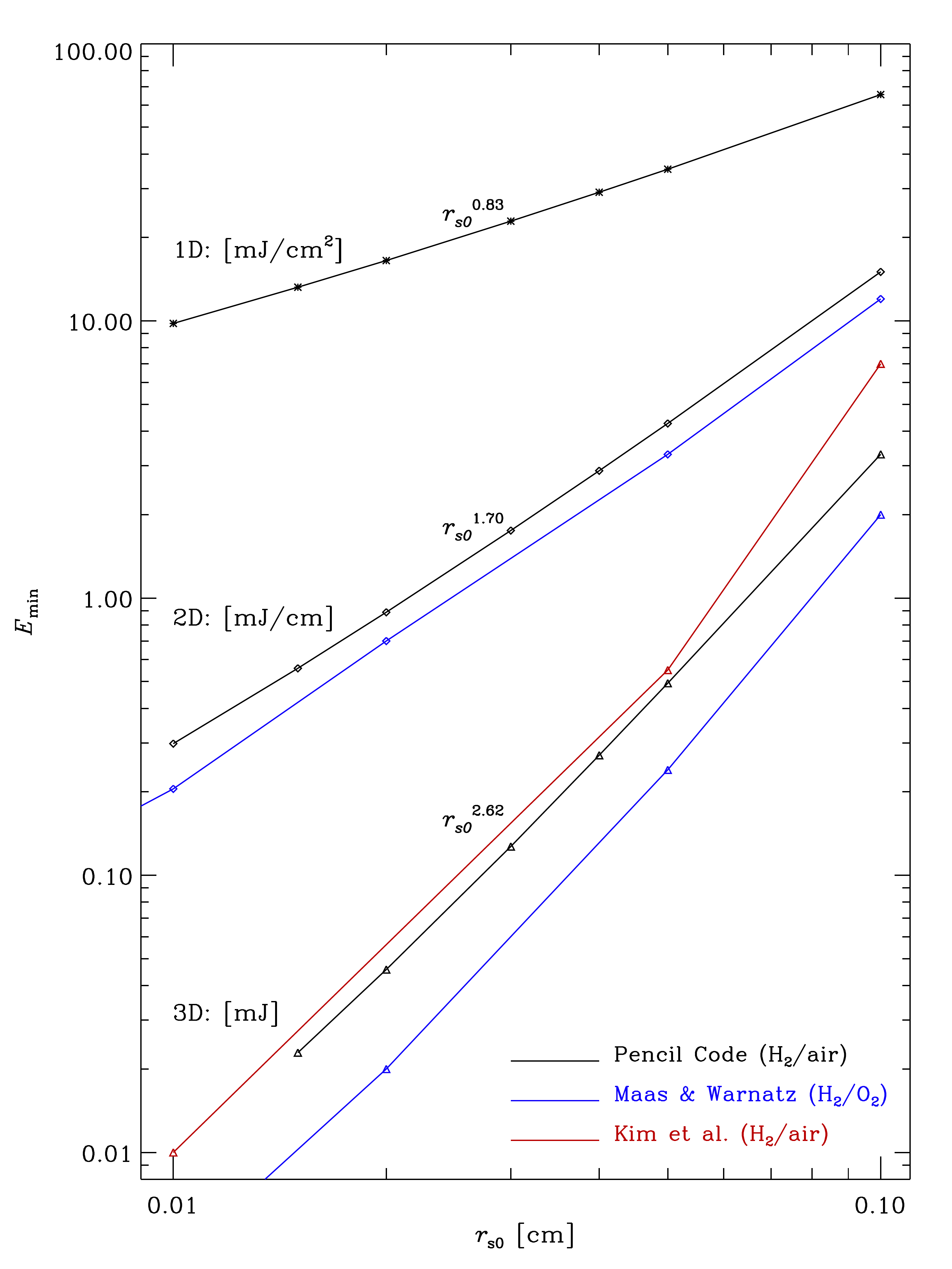}
\caption{MIE as a function of ignition source radius for different 
  dimensionality's. Black line correspond to the DNS results from this work,
blue line are the results for Maas \& Warnatz (1998)~\cite{maas} with a
stoichiometric Hydrogen-Oxygen mixture while the red line is the results of 
Kim et al. (2004)~\cite{kim04} with a stoichiometric Hydrogen-Air mixture.}
\label{FIG: MIE}
\end{figure}
The ignition energy supplied to the mixture is 
\EQ
Q=\int_V c_P\rho (T-T_0)dV'
\EN
where $V$ is a large volume containing the ignition source and $T$ is given by
\Eq{EQ: Gaussian Temp Dist}. 
In order to find MIE for a given $r_s$ a series
of simulations with gradually increasing $T_{\rm max}$ are run. By definition MIE 
equals $Q$ for the lowest temperature $T_{\rm max}$ at which the mixture 
is ignited.
In \Fig{FIG: MIE} the MIE is shown as a function of ignition source radius
for 1, 2 and 3 dimensional setups. For the three dimensional case our DNS
results with the Pencil Code (black line) is slightly above the results of 
Maas \& Warnatz (1998)~\cite{maas}, this is however as expected since 
Maas \& Warnatz (1998)~\cite{maas} considered 
pure Oxygen as oxidizer, while in the current work air has been used. Furthermore,
it is seen that the DNS results are very similar to the results of 
Kim et al. (2004)~\cite{kim04},
which was performed with a stoichiometric Hydrogen-Air mixture, except for the largest radius where the
results of Kim et al. (2004)~\cite{kim04} bends upward. 
It is not known what cause the discrepancy for the largest radii.

For the two dimensional results it is once again seen that the Pencil-Code
results correspond very well with the results of 
Maas \& Warnatz (1998)~\cite{maas} except for 
the vertical shift due to the differences in oxidizer. 

For the one dimensional
line there are no literature results with which it could be compared, but
the trends nevertheless seems to be correct. 

Naively one could think that
the MIE would scale as $r_{s0}^N$ where $N$ is the dimensionality. Due to the
stronger temperature gradients for smaller radii there will however be more
thermal diffusion away from the center of the profile for smaller radii, and
the MIE should therefore scale as $r_{s0}^M$ where $M<N$. This is indeed also
what is found in the simulations where $M$ is 0.83, 1.70 and 2.62, for 1D, 2D and
3D, respectively.

\begin{figure}
\includegraphics[scale=0.4]{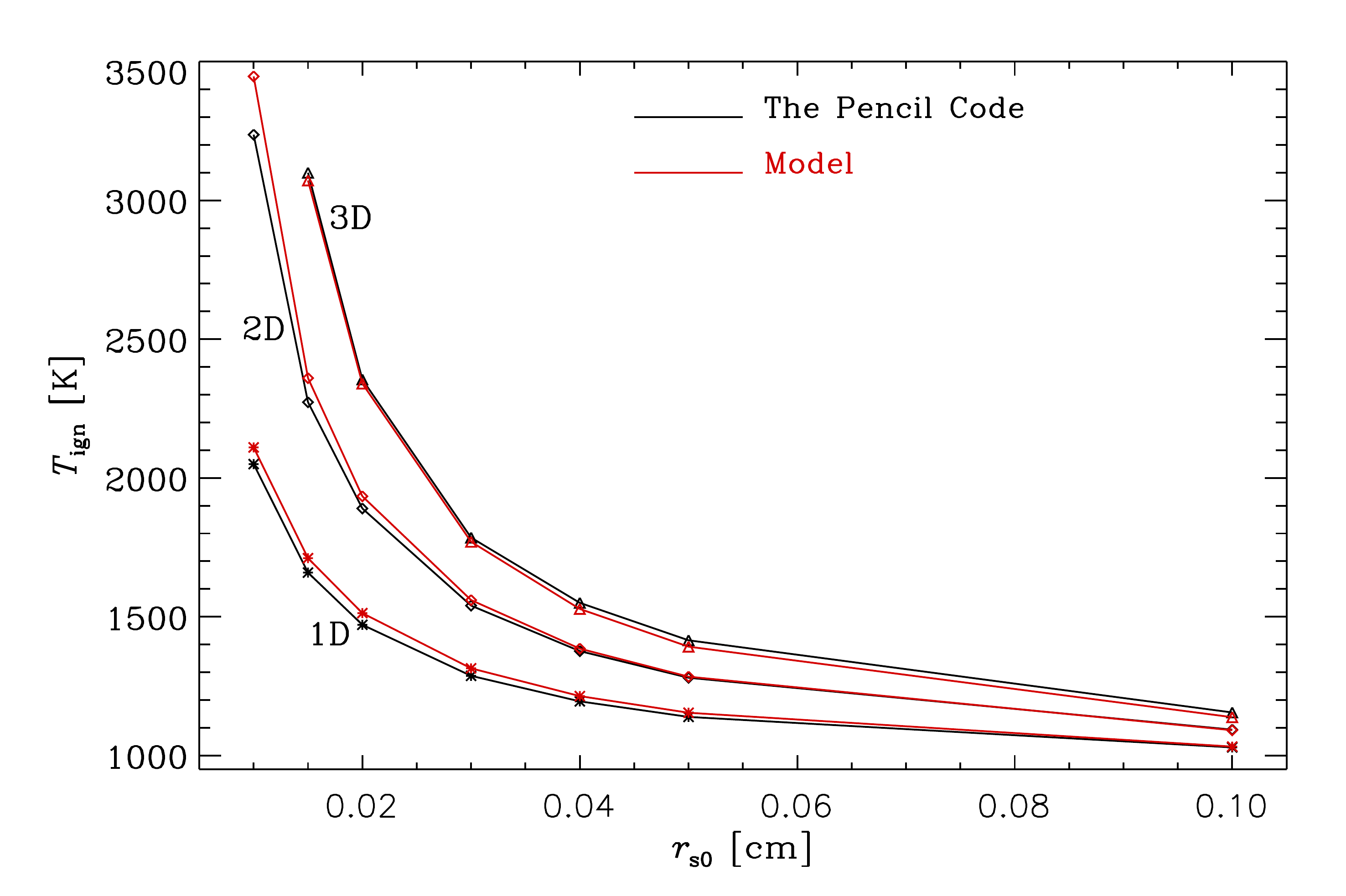}
\caption{In this figure the results from the Pencil Code and the
  results from the simplified model are compared for all the
  dimensions.The parameters are $T_0=300$~K, $\phi=0.8$ and
  $P=1$~atm.}
\label{FIG: Results SMvsPC}
\end{figure}
Lets define an ignition temperature $T_{\rm ign}(r_{s0})$, which is a function of
the initial width of the distribution $r_{s0}$, 
such that for $T_{\rm max}<T_{\rm ign}(r_{s0})$
there is no ignition while for $T_{\rm max}>T_{\rm ign}(r_{s0})$ the mixture 
ignites.
In \Fig{FIG: Results SMvsPC} $T_{\rm ign}$ from the DNS simulations
is shown as a function of $r_{s0}$
for the three different dimensionality's (black line). 
It is clearly seen that as the width of the initial profile
is decreased the required temperature for ignition increases strongly.
This is because the temperature diffusion rate scale as the second order gradient,
which for a Gaussian distribution means that 
$\partial T/\partial t\sim r_s(t)^{-2}$ (see 
\Eq{EQ: 1.order DE Tm(t) with rs(t)}). Furthermore it is seen that
the higher the dimensionality the higher is the required initial temperature.
This comes from the fact that high dimensionality yields more degrees of 
freedom for the thermal diffusion, and consequently the temperature in the
middle of the distribution decreases faster. This is seen in 
\Eq{EQ: 1.order DE Tm(t)} where $\partial T/\partial t\propto N$.

The red lines in \Fig{FIG: Results SMvsPC} show the results obtained with
the new model. It is seen that the model results are very similar to the 
results from the full DNS. Thus it is clear that for the setup used here
the new model is a very good alternative to the full DNS, and there are no
obvious reasons why this should not be the case also for other fuels, 
equivalence ratios, ambient temperatures etc. 

\begin{figure}
\includegraphics[scale=0.4]{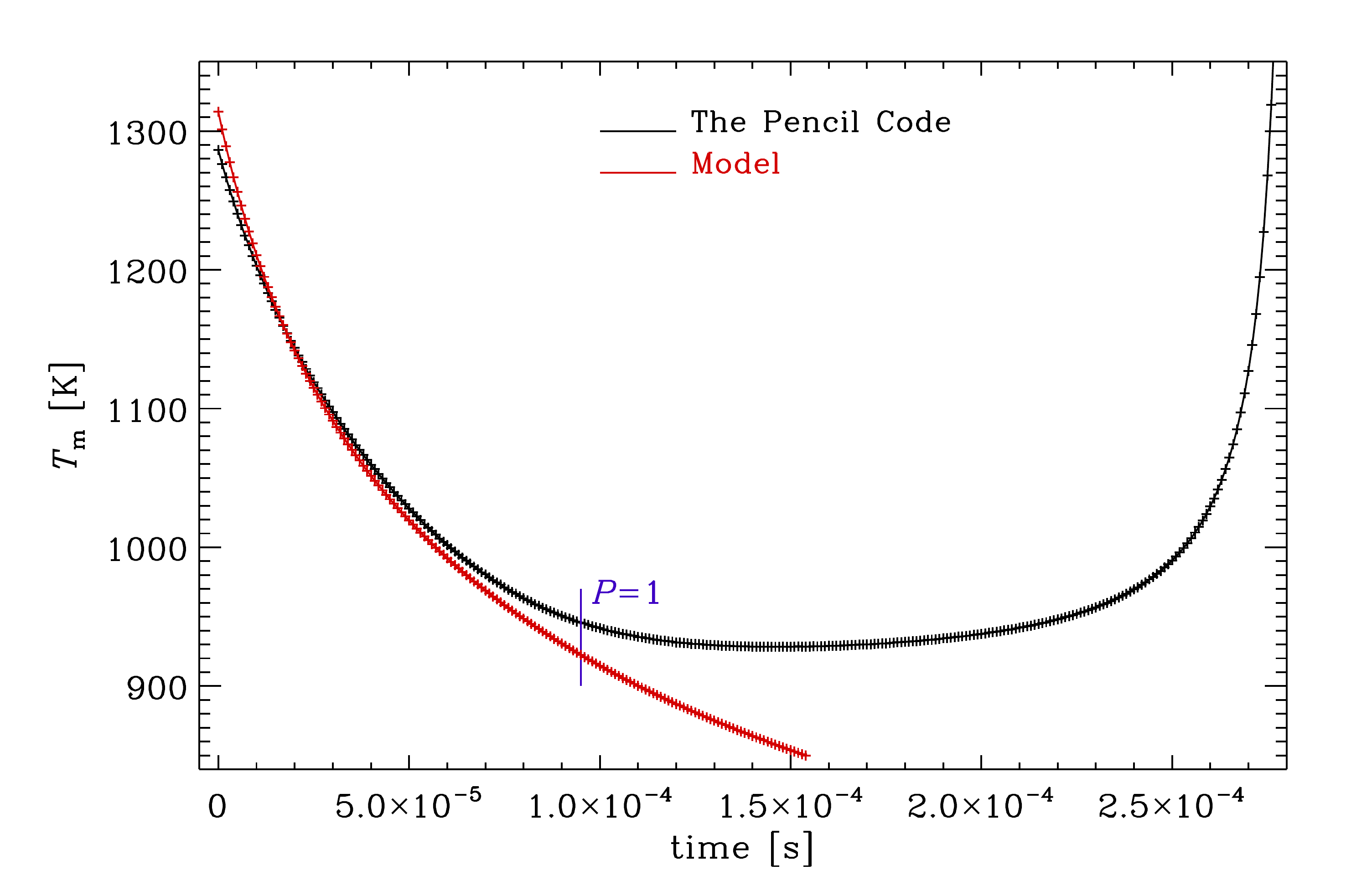}
\caption{Temperature evolution for DNS (black line) and model (red line) results. 
  The time where the ignition progress variable $P$ is unity is marked
  by the blue line. This example is taken from the one dimensional result with $r_{s0}=0.03$~cm.}
\label{FIG: P}
\end{figure}

In \Fig{FIG: P} the 
central temperature evolution is compared for the DNS results and the
model results for the same specific case. The evolution is very similar for early 
times, but they are seen to deviate after $\sim 10^{-4}$~s when
the mixture starts to burn (as we see in the results of DNS)
and thus produce heat. This is however not
a problem since the model calculations will be finalized when $P\geq 1$, 
meaning that ignition has been
achieved, which for this particular case happened after 
$9.5\times 10^{-5}$~s. If, however, the initial temperature is
lower, such that there will be no ignition, the model calculation
will be stopped, with the conclusion that no ignition could be achieved, 
when $P$ has converged at a level below unity.

\section{Conclusion}
A new novel model for determining ignition has been described for the case
of an initial Gaussian temperature distribution. The model, which require
a functional description of the ignition delay time as input, is compared 
against fully resolved DNS results and found to produce very similar results. 
This indicate 
that what determines if an initial ignition kernel will evolve into a 
successful ignition or not relies on two tings; 1) the amount of thermal 
diffusion away from the center of the distribution and 2) the integrated
value of the inverse ignition delay. It is also expected that diffusion of
radicals out of the center of the distribution will have an effect on ignition,
but this is apparently only of equal or less importance compared to the 
thermal diffusion.

Regarding the importance of the dimensionality of the heat source it is found 
that for higher dimensionality's there are more degrees of freedom for the
thermal diffusion, and consequently the temperature will decrease more 
quickly in such a distribution such that the required maximum temperature for
ignition is higher for the higher dimensions.  

\section*{Acknowledgments}
This work was supported by the European 
Community's Seventh Framework Programme (FP7/2007-2013) under grant 
agreement nr 211971 (The DECARBit project).

\end{document}